\documentclass[floatfix, superscriptaddress, preprintnumbers, twocolumn, amsmath, amssymb, aps, pra]{revtex4-2}

\usepackage{graphicx}% Include figure files
\usepackage{dcolumn}% Align table columns on decimal point
\usepackage{bm}% bold math
\usepackage{braket}

\usepackage{xcolor}

\newcommand{\Tr}[1]{\mathrm{Tr} #1}

\begin{document}

\title{Experimental lower bounds to the classical capacity of quantum channels}
\author{Mario A. Ciampini}
\email{mario.arnolfo.ciampini@univie.ac.at}
\affiliation{Vienna Center for Quantum Science and Technology (VCQ), Faculty of Physics,
  University of Vienna, A-1090 Vienna, Austria}
\affiliation{Quantum Optics Group, Dipartimento di Fisica,
  Universit\`a di Roma La Sapienza, Piazzale Aldo Moro 5, I-00185 Roma, Italy}

\author{\'Alvaro Cuevas}
\email{Alvaro.Cuevas@icfo.eu}
\affiliation{ICFO—Institut de Ciències Fot\`oniques,
  The Barcelona Institute of Science and Technology,
  Castelldefels (Barcelona), 08860, Spain}
\affiliation{Quantum Optics Group, Dipartimento di Fisica,
  Universit\`a di Roma La Sapienza, Piazzale Aldo Moro 5, I-00185 Roma, Italy}

\author{Paolo Mataloni}
\email{paolo.mataloni@uniroma1.it}
\affiliation{Quantum Optics Group, Dipartimento di Fisica,
  Universit\`a di Roma La Sapienza, Piazzale Aldo Moro 5, I-00185 Roma, Italy}

\author{Chiara Macchiavello}
\email{chiara@unipv.it}
\affiliation{Dipartimento di Fisica, Universit\`a di
  Pavia, Via Agostino Bassi 6, I-27100 Pavia, Italy}  
\affiliation{INFN-Sezione di Pavia, Via Agostino Bassi 6, I-27100 Pavia, Italy}  
\author{Massimiliano F. Sacchi}
\email{msacchi@unipv.it}
\affiliation{Consiglio Nazionale delle Ricerche - Istituto di
  Fotonica e Nanotecnologie (CNR-IFN), Piazza Leonardo da Vinci 32, I-20133,
  Milano, Italy}
\affiliation{Dipartimento di Fisica, Universit\`a di
  Pavia, Via Agostino Bassi 6, I-27100 Pavia, Italy}

\date{\today}

\begin{abstract}
We show an experimental procedure to certify the classical capacity
for noisy qubit channels. The method makes use of a fixed bipartite
entangled state, where the system qubit is sent to the channel input
and the set of local measurements $\sigma_{x}\otimes\sigma_{x}$,
$\sigma_{y}\otimes\sigma_{y}$ and $\sigma_{z}\otimes\sigma_{z}$ is
performed at the channel output and the ancilla qubit, thus without
resorting to full quantum process tomography. The witness to the
classical capacity is then achieved by reconstructing sets of
conditional probabilities, noise deconvolution, and classical
optimization of the pertaining mutual information. The performance of
the method to provide lower bounds to the classical capacity is tested
by a two-photon polarization entangled state in Pauli channels and
amplitude damping channels.  The measured lower bounds to the channels
are in high agreement with the simulated data, which take into account
both the experimental entanglement fidelity $F=0.979\pm 0.011$ of the
input state and the systematic experimental imperfections.
\end{abstract}

\maketitle
\section{Introduction}
The complete characterization of quantum channels by quantum process
tomography \cite{nielsen97,pcz,mls,dlp,qht,alt,cnot,
  ion,mohseni,irene, atom} is demanding in terms of state preparation
and/or measurement settings, since for increasing dimension $d$ of the
system Hilbert space it scales as $d^4$. When one is interested in
certifying specific properties of a quantum channels, more affordable
procedures can be devised without resorting to complete quantum
process tomography.  This is the case of the detection of
entanglement-breaking properties \cite{qchanndet,qchanndet2} or
non-Markovianity \cite{nomadet} of quantum channels, or the
certification of lower bounds to the quantum capacity of noisy quantum
channels \cite{ms16,ms-corr,exp,correxp}.  Typically, these direct
methods have also the advantage of being more precise with respect to
complete process tomography, which has the drawback of involving
larger statistical errors due to error propagation.

\par One of the most relevant property of quantum channels is the
classical capacity \cite{hol0,sw,hol} for its operational importance
in the quantification of the classical information that can be
reliably transmitted. For the purpose of detecting lower bounds to the
classical capacity, an efficient and versatile procedure has been
recently proposed in Ref. \cite{ms19}.  The method allows to
experimentally detect lower bounds to the classical capacity of
completely unknown quantum channels just by means of few local
measurements, even for high-dimensional systems \cite{multi}.  \par
The gist of the procedure is to efficiently reconstruct a number of
probability transition matrices for suitable input states (playing the
role of ``encoding'') and matched output projective measurements (the
corresponding ``decoding''). The method is accompanied by the
optimization of the prior distribution for the single-letter encoding
pertaining to each input/output transition matrix. In this way, the
mutual information for different communication settings is recovered
and the resulting values are compared. Hence, a lower bound to the
Holevo capacity and then a certification of the minimum reliable
transmission capacity is achieved. Similarly to the method of
certification for the quantum capacity \cite{ms16,ms-corr,exp}, here
each of the conditional probabilities corresponding to a communication
setting can be obtained by preparing just an initial fixed bipartite
state, where only one party enters the quantum channel, while local
measurements are performed at the input and output of the channel.
\par In this paper we present an experimental demonstration for the
above certification of classical capacity in noisy qubit channels. We
implement the method by using highly-pure polarization entangled
photons pairs, where the encoding/decoding settings are achieved by
exploiting complementary observables of the polarization
state. Moreover, a faithful deconvolution of noise is performed over
the experimental data, which allows to optimize the reconstruction of
the probability transition matrices.
\section{Theoretical model}
Let us briefly review the method proposed in Ref. \cite{ms19} for bounding
from below the classical capacity, with specific attention to the case
of single-qubit quantum channels.

The classical capacity $C$ of a noisy quantum channel ${\cal E}$
quantifies the maximum number of bits that can be reliably transmitted
per channel use, and it is given by the regularized expression
\cite{hol0,sw,hol} $C=\lim _{N\rightarrow \infty} \chi ({\cal
  E}^{\otimes N})/N$, with $N$ as the number of uses, in terms of the Holevo
capacity
\begin{eqnarray}
\chi (\Phi) = \max _{\{p_i , \rho _i \}}\{S[\Phi (\textstyle \sum _i p_i \rho
  _i)]-\textstyle 
\sum
_i p_i S[\Phi (\rho _i)]\}
\;,
\end{eqnarray}
where $S(\rho )=-\Tr[\rho \log _2 \rho ]$ denotes the Von Neumann
entropy. The Holevo capacity
$\chi ({\cal E})\equiv C_1$, also known as one-shot classical
capacity,  is a lower bound for the ultimate channel capacity $C$, 
and also an upper bound for the mutual information \cite{mutu,mutu2,mutu3}
\begin{eqnarray}
I(X;Y)=\sum _{x,y}p_x Q(y|x) \log _2 \frac {Q(y|x)}{\sum _{x'} p_{x'} Q(y
  |x')}\;,
\end{eqnarray}
where any transition matrix $Q(y|x)$ corresponds to the conditional
probability for outcome $y$ in an arbitrary measurement at the output
of a single use of the channel with input $\rho _x$, and $p_x$ denotes
an arbitrary prior probability, which corresponds to the distribution
of the encoded alphabet on the quantum states $\{ \rho _x \}$.  \par
As depicted in Fig.~1, in order to obtain a witness for the classical
capacity without resorting to complete process tomography one can
proceed as follows. Prepare a bipartite maximally entangled state
$|\phi^+\rangle=\frac {1}{\sqrt d} \sum_{k=0}^{d-1} |k_{s} \rangle
|k_{a} \rangle $ between a system and ancillary spaces, both with
dimension $d$; send $|\phi ^+\rangle $ through the unknown channel by
keeping the action of ${\cal E}$ on the system alone, namely via the
map ${\cal E}\otimes\mathbb{I} _{a}$; finally, measure locally a
number of observables of the form $X_\alpha \otimes X_\alpha ^{\tau}$,
where $\tau $ denotes the transposition w.r.t. to the fixed basis
defined by $|\phi ^+ \rangle $.

In fact, by denoting the $d$ eigenvectors of $X_\alpha $ as $\{|\phi
^{(\alpha )}_i \rangle
\}$, from the identity \cite{pla}
\begin{eqnarray}
\Tr [(A\otimes B^\tau )({\cal E}\otimes\mathbb{I} _{a})
  |\phi ^+ \rangle \langle \phi ^+ |]
  =\frac 1d   
\Tr[A {\cal E}(B)]\;,\label{pla}
\end{eqnarray}
with arbitrary operators $A$ and $B$, the detection scheme allows to
reconstruct the set of conditional probabilities
\begin{eqnarray}
Q^{(\alpha )}(j|i) = \bra {\phi ^{(\alpha )}_j} 
  {\cal E} (\ket{\phi ^{(\alpha )}_i}\bra{\phi ^{(\alpha
      )}_i})\ket{\phi ^{(\alpha )}_j}  
\;.
\end{eqnarray}
For each encoding-decoding scheme $\alpha $ characterized by the choice of
$X_\alpha $, we can write the corresponding optimal mutual information, namely the
Shannon capacity
\begin{eqnarray}
\!\!\!\!\!\!\!C^{(\alpha )}\!=\!\max _ {\{p_i^{(\alpha )} \}} \sum _{i,j}p_i^{(\alpha )} 
Q^{(\alpha )}(j|i)  
\log _2 \frac{Q^{(\alpha )}(j|i)  }{\sum _l 
p_l^{(\alpha )} Q^{(\alpha )}(j|l) }
\,.\label{ii}
\end{eqnarray}
\begin{figure}[h]
    \includegraphics[width=0.4\textwidth]{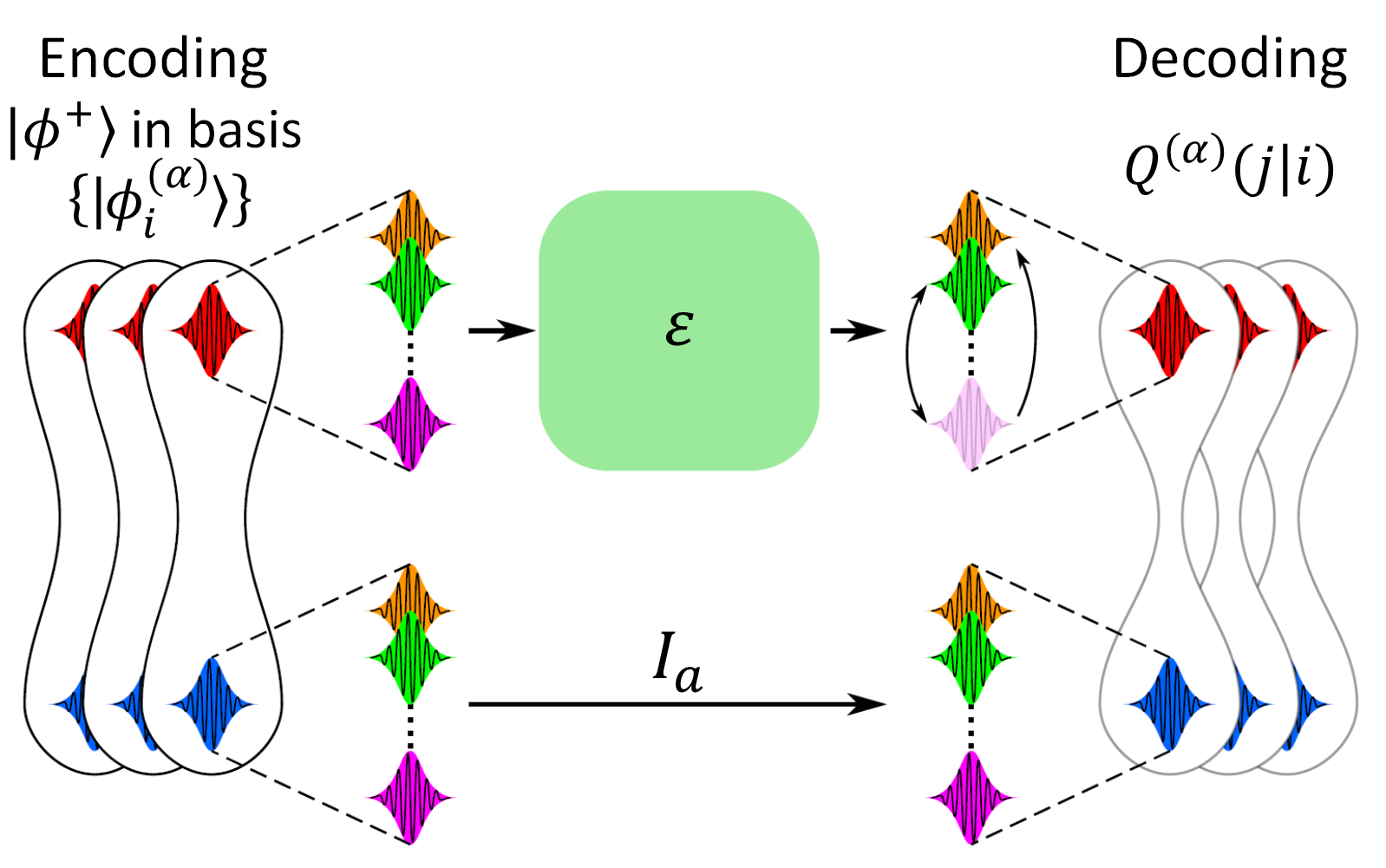}
    \caption{Protocol's concept: The $s$-qudit of a maximally
      entangled bipartite quantum state $\ket{\phi^{+}}$ is sent
      through a noisy quantum channel ${\cal E}$, while the $a$-qudit
      propagates freely. Encoding-decoding of type $\alpha $ is
      represented by layers and is achieved by measuring locally the
      basis $\{\ket{\phi_{i}^{(\alpha )}}\}$. The conditional
      probability $Q^{(\alpha )}(j|i)$ of measuring
      $\ket{\phi_{j}^{(\alpha )}}$ for an input
      $\ket{\phi_{i}^{(\alpha )}}$ allows to evaluate the Shannon
      capacity $C^{(\alpha )}$. The witness $C_D$ of the classical
      capacity is the largest of the values $\{C^{(\alpha )}\}$ among
      all tested types $\alpha $.}
    \label{fig:concept}
\end{figure}
Then, we have the chain of inequalities
\begin{eqnarray}
C 
\geq C_1 \geq C_{D} \equiv \max _\alpha \ \{C^{(\alpha )}\}
\;,
\end{eqnarray}
where $C_{D}$ is the experimentally accessible witness that depends on
the chosen set of measured observables labeled by $\alpha $, and
provides a lower bound to the classical capacity of the unknown
channel.
\par In the present scenario we consider single-qubit channels 
and the information settings correspond to the choice of
the three local observables $\sigma_{x}\otimes\sigma_{x}$,
$\sigma_{y}\otimes\sigma_{y}$ and
$\sigma_{z}\otimes\sigma_{z}$. Hence, each
of the three conditional probabilities $Q^{(\alpha)}(j|i)$, with
$\alpha =x,y,z$, is a $2\times 2$ transition matrix that corresponds to a
binary classical channel, for which the optimal prior probability
$\{p_i ^{(\alpha)}\}$ can be theoretically evaluated \cite{ms19}. In
fact, without loss of generality, for each transition matrix we can
fix the labeling of logical zeros and ones such that $0\leq \epsilon _0\leq
\frac 12$, $\epsilon _0 \leq \epsilon _1$, and $\epsilon _0 \leq 1 -
\epsilon _1$, where $\epsilon _0 $ denotes the error probability of
receiving $1$ for input $0$, and $\epsilon _1 $ denotes the error
probability of receiving $0$ for input $1$.  Then, for each $\alpha
=x,y,z$ the mutual information as in Eq. (\ref{ii}) is maximized by
a prior probability $\{p_0,p_1=1-p_0\}$, with
\begin{eqnarray}
p_0= \frac{1-\epsilon _1 (1+z)}{(1-\epsilon _0-\epsilon _1)(1+z)}\;,\label{po}
\end{eqnarray}
where $z=2^{\frac {H[\epsilon _0]-H[\epsilon _1]}{1-\epsilon _0 -\epsilon
    _1}}$, and $H(p)=-p \log _2 p -(1-p)\log_2 (1-p)$ denotes the binary Shannon
entropy. The corresponding capacity is
given by \cite{ms19}
\begin{eqnarray}
&&  C_B(\epsilon _0, \epsilon _1)
  =\log _2\left[ 1+ 
2^{\frac {H[\epsilon _0]-H[\epsilon _1]}{1-\epsilon _0 -\epsilon
    _1}}\right ] \nonumber \\& &
  +\frac{\epsilon _0}{1-\epsilon _0 -\epsilon
  _1}H[\epsilon _1] -\frac{1-\epsilon _1}{1-\epsilon _0 -\epsilon
  _1}H[\epsilon _0]
\;.\label{c01} 
\end{eqnarray}
Notice that for $\epsilon _0=\epsilon _1 =\epsilon $ one
recovers the classical capacity for the binary symmetric channel 
\begin{eqnarray}
C_B(\epsilon , \epsilon )= 1-H[\epsilon ]\;,
\end{eqnarray}
with uniform optimal prior $\{1/2,1/2 \}$,
whereas for $\epsilon _0=0$
(i.e., when  only input 1 is affected by error) one obtains the capacity of
the so-called $Z$-channel
\begin{eqnarray}
C_B(0,\epsilon )=\log _2 [1 + (1-\epsilon ) \epsilon ^{\frac {\epsilon
    }{1-\epsilon }}]\;.\label{cb0}
\end{eqnarray}
In Appendix~\ref{app:A} we summarize the expected
theoretical results for the qubit channels that we implemented experimentally.

In order to consider the experimental imperfections in our state
preparation, we assume a Werner form
\begin{eqnarray}\label{werner}
\rho_{F}=\frac{4F-1}{3}
\ket{\Phi^{+}}\bra{\Phi^{+}}+\frac{1-F}{3}\mathbb{I}_s\otimes \mathbb{I}_{a}
  \;,\label{rhof}
\end{eqnarray}
where $F=\braket{\Phi^{+}|\rho _{F}|\Phi^{+}}$ denotes the fidelity with
respect to the ideal maximally entangled state $\ket{\Phi^{+}}
= \frac{1}{\sqrt 2} \left(\ket{0}\ket{0}+\ket{1}\ket{1}\right)$.
  Hence, we can replace Eq. (\ref{pla}) with
\begin{eqnarray}
&&  \Tr [A {\cal E}(B)]= \frac{2}{4F-1} \times  \label{pla2}  \\
&& 
  \Tr \{[ A \otimes (3 B^\tau
    -  2 (1-F)\Tr[B] \mathbb{I}_a)]
  ({\cal E}\otimes \mathbb{I}_a) \rho _F \}
  \,.\nonumber 
\end{eqnarray}
Equation (\ref{pla2}) allows one to deconvolve the noise as long as
$F\neq1/4 $, since the output state $({\cal E}\otimes \mathbb{I}_{a})
\rho _F $ faithfully represents the unknown channel ${\cal E}$.  \par
The normalized conditional probabilities for the three binary
classical channels corresponding to the information settings $\alpha
=x,y,z$ where coding and decoding are performed using the eigenstates
of the three Pauli matrices $\sigma _x, \sigma _y,\sigma _z $ can be
obtained by the ratio $\frac{\Tr [A {\cal E}(B)]}{\Tr [{\cal E}(B)]}$,
when replacing $A$ and $B$ with the pertaining eigenvectors. In
practical terms, these quantities are measured from the bipartite
qubit detections, which in optical channels corresponds to two-photon
coincidence detections $C(ji)=
\Tr[(|j \rangle \langle j|_S \otimes |i
  \rangle \langle i |_A ) ({\cal E} \otimes \mathbb{I}_a) \rho _F ]$
on the output state. We provide in
Appendix~\ref{app:B} the explicit expressions for the conditional
probabilities in terms of the measured coincidences. For each
information setting $\alpha =x,y,z$, the logical bits and their
pertaining transition error
probabilities $\epsilon ^{(\alpha )}_0$ and $\epsilon ^{(\alpha )}_1$
introduced before Eq. (7) are identified from the conditional
probabilities as follows
\begin{eqnarray}
  \!\!\!\!\!\!\!\!\!
  \epsilon ^{(\alpha )}_0 =\min _{\{i,j\}} Q^{(\alpha )} (i|j) \equiv Q^{(\alpha)}(1|0)\,;
\ \   \epsilon ^{(\alpha )}_1 \equiv Q^{(\alpha )}(0|1) 
  \,.\label{dett}
\end{eqnarray}
Using Eq. (\ref{c01}), from the experimental data we then obtain
$C^{(\alpha )}=C_B(\epsilon ^{(\alpha )}_0, \epsilon
^{(\alpha )}_1)$, and hence the detected classical capacity as
\begin{eqnarray}
C_{D}= \max _{\alpha =x,y,z} C_B(\epsilon ^{(\alpha )}_0, \epsilon
^{(\alpha )}_1)\;.
\end{eqnarray}
We remark that the effect of a fidelity value $F<1$ (except the case
$F=\frac 14$) is not detrimental for accessing the capacity witness
$C_{D}$, and only the statistical noise is expected to increase for
decreasing value of $F$. This can be argued from the identity
(\ref{pla2}) which, for any couple of bases $\{|i \rangle _A
\}_{i=0,1}$ and $\{ |j \rangle _S \}_{j=0,1}$, provides an unbiased
estimation of the joint probabilities $p_{F}(j,i)=\Tr[(|j \rangle
  \langle j |_S \otimes |i \rangle \langle i |_A ) ({\cal E} \otimes
  \mathbb{I}_a) \rho _{F}]$ pertaining to the case of ideal fidelity
$F=1$ in terms of the measured probabilities as
\begin{eqnarray}
  \!\!\!\!\!\!\!\!\!
  p_{F=1}(j,i)&& =\frac{1+2F}{4F-1} p_F (j,i)  
  -  \frac {2(1-F)}{4F-1}
  p_F (j,i \oplus 1) \,.
\end{eqnarray}
By denoting with $\sigma $ the typical order of magnitude of the
standard deviation for the measured joint probabilities with $F<1$,
then the noise deconvolution provides the correct probabilities with
standard deviation
\begin{eqnarray}
\sigma _F && = \sigma \sqrt{\left( \frac{1+2F}{4F-1}\right )^2 +
  4\left (\frac{1-F}{4F-1} \right )^2}\nonumber \\& &=
\sigma \frac{\sqrt{8F^2-4F+5}}{|4F-1|}\;.\label{std}
\end{eqnarray} 
\section{Experimental implementation}
\begin{figure}[t]
	\centering
    a)\includegraphics[width=0.48\textwidth]{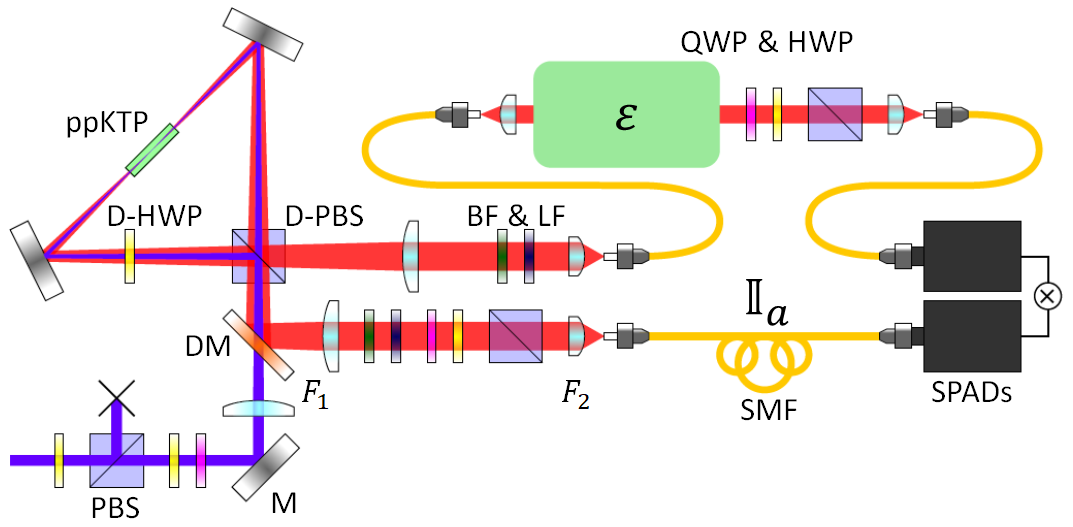}\\
    b)\includegraphics[width=0.4\textwidth]{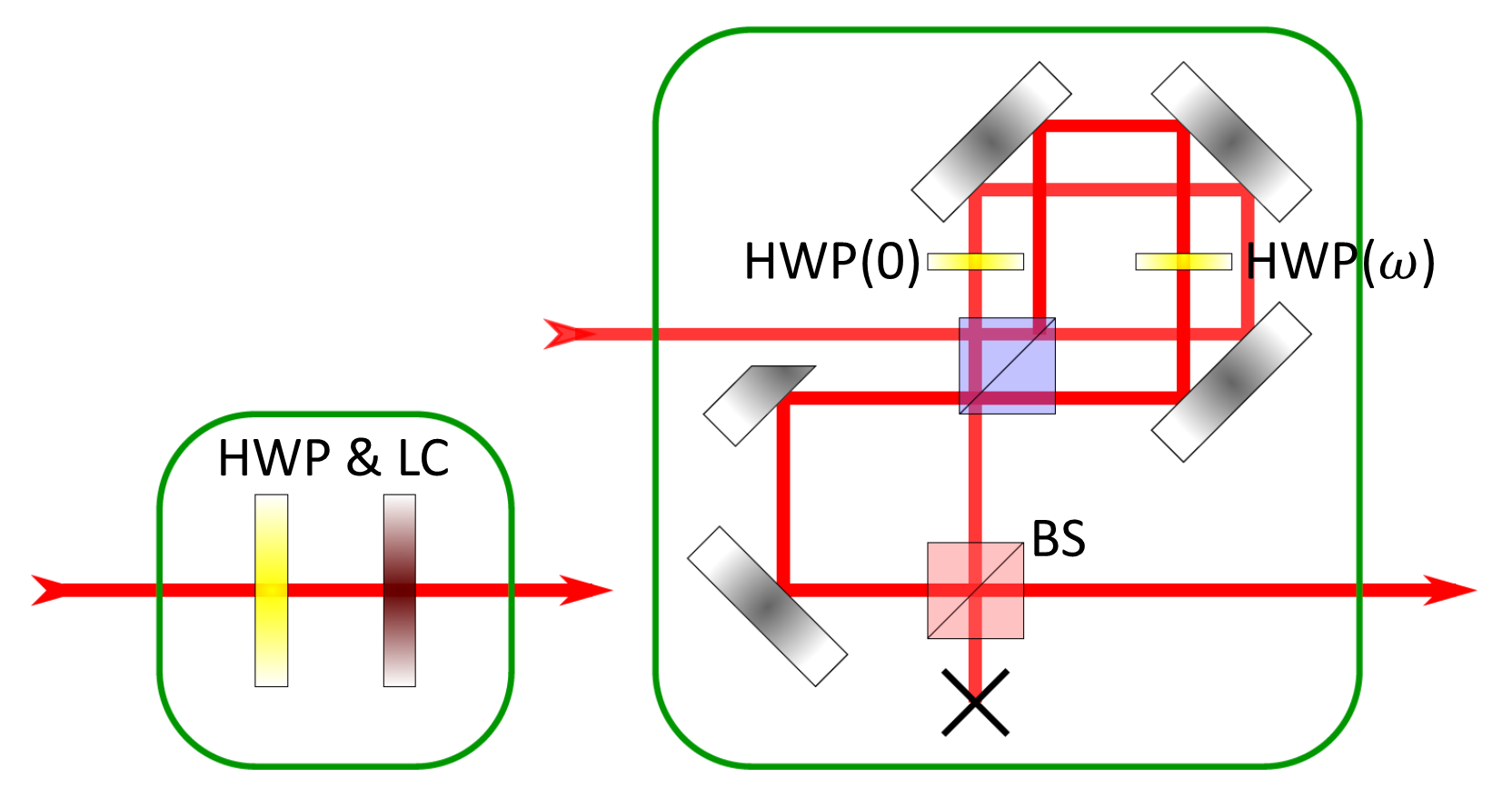}
	\caption{a) Full scheme with quantum source, propagation
          channels, projection and measurement devices. PBS:
          polarizing beam splitter; M: mirror; BS: non-polarizing
          50:50 beam splitter; QWP: quarter wave-plate; HWP: half
          wave-plate; DM: dicroic mirror; ppKTP: periodically poled
          non-linear crystal; BF: band-pass filter; LF: long-pass
          filter; $F_{k}$: lenses; SPADs: single photon avalanche
          detectors; SMF: single mode fiber; LC: liquid
          crystal element. Additional D-label for compatibility with
          both pump and generated wavelength. b) Left: Setup for
          phase-damping and depolarizing channels. Right: Setup for
          amplitude damping channel.}
	\label{fig:setup}
\end{figure}

A single-mode continuous-wave laser at 405nm was used to pump a
Type-II ppKTP crystal within a Sagnac interferometer (SI) in order to
produce polarization-entangled photon pairs at 810nm via spontaneous
parametric down conversion, as implemented in \cite{exp}. The two
output modes were labeled as $s$-qubit and $a$-qubit, so the Bell
state
$\ket{\Phi^{+}}=\frac{1}{\sqrt{2}}\left(\ket{H_{s}}\ket{H_{a}}+\ket{V_{s}}\ket{V_{a}}\right)$
was generated. As shown in the full scheme of Fig.~\ref{fig:setup},
tube lenses with $F_{1}=300mm$ collimate both output modes, while
objective lenses with $F_{2}=11mm$ couple them into single-mode-fibers
(SMFs).

The fidelity of the experimental state $\rho_{F}$ defined in
Eq.~(\ref{werner}) and measured by quantum tomography techniques
\cite{tomo1, tomo2} was
$F=\braket{\Phi^{+}|\rho_{F}|\Phi^{+}}=0.979\pm 0.011$. Hence, 
we use Eq.~(\ref{pla2}) for noise deconvolution, and the standard
deviation of the joint probabilities are increased just by $1.45\%$ with
respect to the ideal case $F=1$, according to Eq. (\ref{std}).

The observables $\sigma _{\alpha }$, for $\alpha =x,y,z$, are measured
with rotations of a quarter wave-plate (QWP) and a half wave-plate
(HWP) before a polarizing beam splitter (PBS) for each
qubit. Remaining photon pairs after these projections are measured
with synchronized single-photon avalanche detectors (SPADs) within a
time window of 5ns for an integration time of 10s/measurement.
\par By referring to the parametrization of channels given in
Appendix~\ref{app:A}, within the family of Pauli channels we have
chosen two special cases as implemented in \cite{exp,pc}. The first
one is the phase damping (PD) channel (described by $q_z=q$ and
$q_{x}=q_{y}=0$ in Appendix~\ref{app:A}). The second one is the
depolarizing (D) channel (described by $q_x=q_y=q_z=q$). Their
experimental simulations were implemented by $q_{i}$-weighted
combinations of independent Pauli operations, achieved by setting a
HWP at $0^{\circ}$-degrees ($45^{\circ}$-degrees) for $\sigma_{z}$
($\sigma_{x}$) and a liquid crystal (LC) at minimum (maximum) voltage
for $\sigma_{z}$ ($\mathbb{I}$) [see
  Fig.~\ref{fig:setup}b]. Accordingly, the simultaneous operation
$\sigma_{y}=i\sigma_{x}\sigma_{z}$ give us the last Pauli operation.

\par The experimental simulation of an amplitude damping (AD) channel
was achieved by using a square SI with displaced trajectories as
implemented in \cite{exp,adc2}, where a HWP at $0^{\circ}$-degrees was
placed in the H-polarized counter-clockwise trajectory, while another
HWP at $\omega=\frac{\arccos(-\sqrt{1-\eta})}{2}$ degrees was place in
the V-polarized clock-wise trajectory. The induced rotation of a
single polarization (or damping) sends this light to the long arm of
an unbalanced Mach-Zehnder interferometer (MZI), which then
re-combines it out of coherence in a non-polarizing beam splitter
(BS).

\section{Results}
\begin{figure*}[ht!]
	\centering
    \includegraphics[width=.98\linewidth]{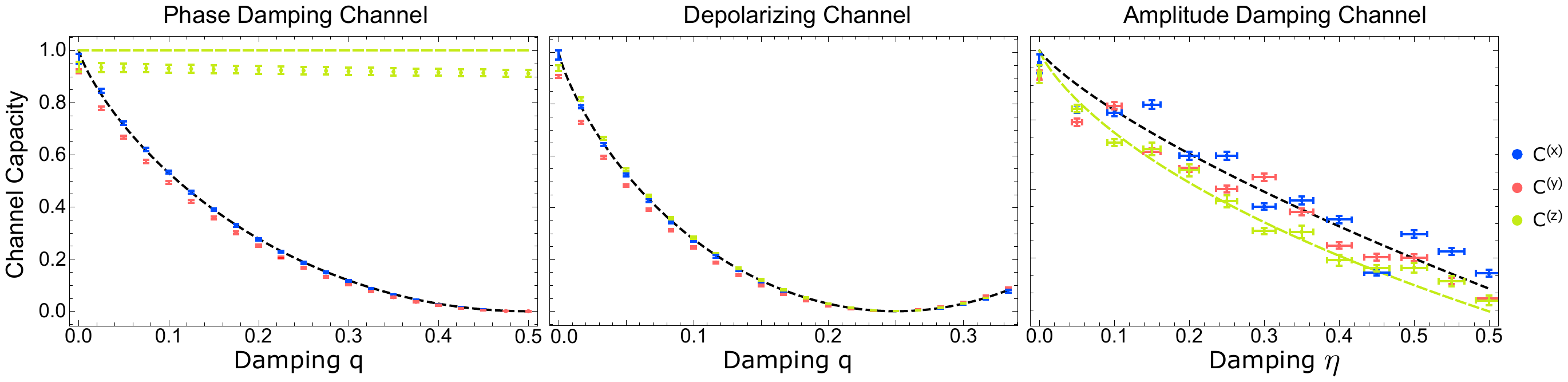}
	\caption{\textbf{Experimental Channel Capacity.} Left:
          $C^{(\alpha )}$ for PD-channel. Center: $C^{(\alpha )}$ for
          D-channel. Right: $C^{(\alpha )}$ for AD-channel.  In blue
          (dark grey) $C^{(x)}$, in red (grey) $C^{(y)}$, in green
          (light grey) $C^{(z)}$. For the predicted theoretical
          behavior, $C^{(z)}$ is depicted with dashed green (light
          grey), while $C^{(x)}$ and $C^{(y)}$ with dashed black
          lines. All error-bars (standard deviation) are calculated
          from at least 8 values per point; their $x$-axis components
          originate from the propagated uncertainty in the angle of
          the HWPs, which dominates over other sources of error, like
          wrong retardance in the LC.}
	\label{fig:results}
\end{figure*}

In Fig.~\ref{fig:results} we show the measured Shannon capacities
$C^{(\alpha )}$ in all three bases for the experimental simulation of
three different environments, namely a PD-channel, a D-channel, and an
AD-channel. This is compared to the predicted behaviour for the real
input state, which is fidelity-dependent from Eq.~(12) and in detail
explained in the Appendix~\ref{app:B}. See Appendix~\ref{app:C} for
details on the noise propagation and Appendix~\ref{app:D} for details
on the experimental raw data.
\par For a PD-channel the witness $C_D$ of the channel capacity is obtained
for $\alpha =z$ and can be compared with the expected theoretical
value $C_D=1$. The presented results confirm the behavior in which
phase-damping processes do not affect binary data transmission in the
logical basis, but only in complementary superposition bases. The
systematic offset below unity value is due to the propagation of error
by replacing experimental negative values of $\epsilon _0^{(z)}$ with
their absolute values.  We observe that the customary constrained
maximum-likelihood technique can solve the possible issue of negative
reconstructed probabilities and the resulting bias in the capacity
bounds.

For a D-channel the theoretical value $C_D=1-H(2q)$ is equally
achieved by any of the three bases. Here, we find the best agreement
with respect to this prediction, because small differences of the
experimental state or channel from the ideal ones are rapidly
compensated during the decoherence process by the simultaneous action
of all three Pauli operations. Accordingly, the balance increase
(decrease) in the diagonal (off-diagonal) terms of the density matrix
reduces bias projections and final data dispersion.

For an AD-channel the experimental implementation requires control on
multiple optical paths, where systematic error can have unbalanced
contribution among different bases. Such an issue, combined with the
non-linear dependency of $\eta$ with respect to $\omega$ and an
angular uncertainty of $0.5^{\circ}$-degrees, gives considerable
propagated errors. The expected capacity witness $C_D=1-H\left (
\frac{1-\sqrt{1-\eta }}{2} \right )$ is achieved equivalently by the
bases $x$ or $y$. Our method still succeeds in showing the general
behavior and providing a sensible lower bound to the classical channel
capacity.

In summary, by comparison with the theoretical predictions, the method
proved to be very effective in providing a witness for the classical
capacity of noisy quantum channels, even if the channel implementation
has critical and systematic imperfections. Moreover, by means of the noise
deconvolution, we have shown that the method is robust with respect to
an imperfect probe-state preparation and provides an unbiased estimation
of the theoretical lower bounds.  

\section{Conclusions} 
The experimental coincidence counts allow to directly reconstruct the
sets of conditional probabilities for three different information
settings. For each setting $\alpha =x,y,z$, by Eq.~(\ref{dett}) the
logical bits and the values of $\epsilon _0^{(\alpha )}$ and $\epsilon
_1^{(\alpha )}$ are identified. From these values one recovers three
values of $C_B$ as in Eq.~(\ref{c01}), which correspond to the Shannon
capacity, namely the optimized mutual information for coding/decoding
on the eigenstates of $\sigma _x$, $\sigma _y$, and $\sigma _z$. The
highest among such three values is selected as a classical capacity
witness, namely it certifies a lower bound to the ultimate classical
capacity of the unknown quantum channel. For each information setting,
by Eq.~(\ref{po}) the identified values of $\epsilon _0^{(\alpha )}$
and $\epsilon _1^{(\alpha )}$ also allow one to obtain the optimal
weight for the coding of the two logical characters.

We emphasize that our method is highly robust against experimental
imperfections, even in cases where the prepared quantum state is
partially mixed as the Werner state. For each of the three chosen
bases, the Shannon capacity is precisely retrieved for all values of
the channel parameters. Nevertheless, the most accurate values were
obtained for the Pauli channels, which are effectively produced by
mixing unitary operations over the input state. For mixtures of
non-unitary operations as the amplitude damping channel, we find a
less smooth behavior in the detected classical capacity because
experimental imperfections do not act in balanced ways for both photon
polarizations.

We recommend this novel method for certifying the classical capacity
of quantum channels due to its high precision under different
scenarios. The studied cases were not particularly designed to match
with the protocol, but chosen to represent a variety of classes. No
prior information is needed about the channel structure, while the
measurement settings are much less demanding with respect to full
process tomography.

\begin{acknowledgments}
{\em Acknowledgments.}  \'A.C. work was supported by Becas Chile
N$^{\circ}$74200052 from the Chilean National Agency for Research and
Development (ANID). C.M. acknowledges support by the European Quantera
project QuICHE.
\end{acknowledgments}

\appendix
\section{Expected theoretical results for the implemented qubit channels}\label{app:A}
For an \textbf{Amplitude Damping (AD) channel}, described by
\begin{eqnarray}
{\cal E}^{AD}(\rho )= K_0 \rho K_0^\dag + K_1  \rho K_1^\dag \;, 
\end{eqnarray}
where $K_0= |0 \rangle \langle 0| + \sqrt {1- \eta }|1 \rangle
\langle 1|$ and $K_1= \sqrt \eta |0 \rangle \langle 1|$, the error
probabilities of the three binary channels correspond to
\begin{eqnarray}
&&\epsilon _0^{(x)}=\epsilon _1^{(x)}=
  \epsilon _0^{(y)}=\epsilon _1^{(y)}=(1 - \sqrt{1-\eta })/2\,;\nonumber\\
&& \epsilon _0^{(z)}=0\,; \quad \epsilon _1^ {(z)}=\eta \,. 
\end{eqnarray}
The detected capacity is obtained equivalently by the $x$ or $y$
information setting and is given by
\begin{eqnarray}
  C_{D}=C^{(x)}=C^{(y)}=1- H\left ( \frac {1-\sqrt{1-\eta}}{2}
  \right )
  \;.\label{unc}
\end{eqnarray}
For any $\eta $, this result outperforms the $z$-setting, for which
one has $C^{(z)}=C_B(0,\eta )$, according to Eq. (10).
We recall that the detected capacity in Eq.~(\ref{unc}) is strictly
lower than the Holevo capacity $C_1$, which can be evaluated as
\cite{gf,gf2}
\begin{eqnarray}
C_1 = \max _{t\in [0,1]} \left (
 H[t (1 - \eta ) ] - H\left [g(\eta ,t)
\right ]\right )
\;,\label{c1g}
\end{eqnarray}  
where $g(\eta ,t)\equiv\frac 12 [1 + \sqrt{1 - 4 \eta (1 - \eta )
t^2}]$. \\

For a \textbf{Pauli (P) channel}, described by
\begin{equation}
    {\cal E}^{P}(\rho )=q_{\mathbb{I}}\rho + q_x \sigma
_x \rho \sigma _x + q_y \sigma _y \rho \sigma _y +q_z \sigma _z \rho \sigma _z\,,
\end{equation} 
with $q_{\mathbb{I}}=1-(q_{x}+q_{y}+q_{z})$ and 
$q_{x}+q_{y}+q_{z}\leq 1$ (with $q_i\geq 0$), the three channels that are
compared are symmetric and correspond to the error probabilities
\begin{eqnarray}
&&\epsilon _0^{(x)}=\epsilon _1^{(x)}=\min \{q_{y}+q_{z},1-(q_y+q_z)\};\nonumber \\
& & \epsilon _0^{(y)}=\epsilon _1^{(y)}=\min \{q_{x}+q_z,1-(q_x+q_z)\}; \\
& &\epsilon _0^{(z)}=\epsilon_1^{(z)}=\min\{q_{x}+q_{y},1-(q_{x}+q_{y})\}.\nonumber
\end{eqnarray}
The detected capacity is then given by 
\begin{equation}\label{pauli_unc}
C_{D}=1-\min\{H(q_y+q_z),H(q_x+p_z),H(q_x+p_y)\},
\end{equation}
where $\min\{\cdot\}$ compares the $x$, $y$, and $z$ information
settings, respectively. We recall that for Pauli channels the capacity
witness $C_{D}$ equals the Holevo and the classical capacity
(i.e., $C=C_1=C_{D}$, since the additivity hypothesis holds true for
unital qubit channels \cite{unital}).
\newpage
\begin{widetext}
\section{Probability transition matrices for the three information
  settings $\sigma _z$, $\sigma _x$, $\sigma _y$}\label{app:B}
From Eq. (12) the ratio $ \frac{\Tr [A {\cal E}(B)]}{\Tr [{\cal
      E}(B)]}$ is given by
\begin{eqnarray}\label{cond}
%  &&
  \frac{\Tr [A {\cal E}(B)]}{\Tr [{\cal E}(B)]}=
%  \nonumber \\& & 
  \frac{\Tr \{[ A \otimes (3  B^\tau   -  2
      (1-F)\Tr[B] \mathbb{I}_a)] ({\cal E}
    \otimes \mathbb{I}_{a})
    \rho _F \}}{
  \Tr \{[ \mathbb{I}_s \otimes ( 3 B^\tau  - 2
    (1-F)\Tr[B] \mathbb{I}_{a})] ({\cal E}
  \otimes \mathbb{I}_{a})
  \rho _F \}}\,.\label{ratio}
\end{eqnarray}
When qubits are encoded in the single-photon polarization degree of
freedom, the $z$-information setting corresponds to choice of $A$ and
$B$ as the projectors on the basis $\{\ket{H},\ket{V}\}$, with $H$
($V$) as the horizontal (vertical) polarization. Using
Eq. (\ref{ratio}) the corresponding probability transition matrix is
then given in terms of the measured coincidences $C(ji)=\Tr[(|j
  \rangle \langle j|_S \otimes |i \rangle \langle i |_A ) ({\cal E}
  \otimes \mathbb{I}_a) \rho _F ]$ as
\begin{eqnarray}
&& Q^{(z)}(H|H) = \frac{(1+2F)C(HH) -2(1-F)C(HV)}{(1+2F)[C(HH)
      +C(VH)]-2(1-F)[C(HV)+C(VV)]},
\qquad   Q^{(z)}(V|H) = 1- Q^{(z)}(H|H),\\
&& Q^{(z)}(H|V) = \frac{(1+2F)C(HV)
  -2(1-F)C(HH)}{(1+2F)[C(HV)+C(VV)]-2(1-F)[C(HH)+C(VH)]},
\qquad Q^{(z)}(V|V) = 1- Q^{(z)}(H|V).
\end{eqnarray}
For the $x$-information setting $A$ and $B$ correspond to the projectors on the
diagonal basis $\ket{+}=\frac{1}{\sqrt 2}(|H \rangle +|V \rangle ) $ and 
$|- \rangle=\frac{1}{\sqrt 2}(|H \rangle -|V \rangle ) $, and so one has
\begin{eqnarray}
&& Q^{(x)}(+|+) =   \frac{(1+2F)C(++) -2(1-F)C(+-)}{(1+2F)[C(++)+C(-+)]-2(1-F)[C(+-)+C(--)]},
\qquad   Q^{(x)}(-|+) = 1- Q^{(x)}(+|+),\\
&& Q^{(x)}(+|-) = \frac{(1+2F)C(+-)
  -2(1-F)C(++)}{(1+2F)[C(+-)+C(--)]-2(1-F)[C(++)+C(-+)]},
\qquad Q^{(x)}(-|-) = 1- Q^{(x)}(+|-).
\end{eqnarray}
Finally, for the $y$-information setting $A$ and $B$ correspond to
the projectors on the circular basis $|R \rangle =\frac{1}{\sqrt 2}(|H
\rangle -i |V \rangle ) $ and $|L\rangle=\frac{1}{\sqrt 2}(|H \rangle
+ i |V \rangle )$, and hence
\begin{eqnarray}
&&  Q^{(y)}(R|R) = \frac{(1+2F)C(RL)
    -2(1-F)C(RR)}{(1+2F)[C(RL)+C(LL)]-2(1-F)[C(RR)+C(LR)]},
\qquad  Q^{(y)}(L|R) = 1- Q^{(y)}(R|R),\\
&& Q^{(y)}(R|L) = \frac{(1+2F)C(RR)
  -2(1-F)C(RL)}{(1+2F)[C(RR)+C(LR)]-2(1-F)[C(RL)+C(LL)]},
\qquad Q^{(y)}(L|L) = 1- Q^{(y)}(R|L).
\end{eqnarray}
\end{widetext}
Notice the different symmetry in the equations for the $y$-coding with
respect to the $z$ and $x$ cases, due to the presence of the
transposition in Eq. (\ref{ratio}).  For each of the above binary
classical channels the transition errors $\epsilon ^{(\alpha )}_0$ and
$\epsilon ^{(\alpha )}_1$ are identified from the conditional
probabilities by Eq. (13).

According to the above equations, all expectation values are obtained
just by the measurement of 12 polarization projections of the state (4
by each of the 3 observables), which makes the process efficient in
terms of registration and analysis of data, given the reduced number
of operations compared with standard process tomography.

\section{Detection efficiencies and experimental error analysis}\label{app:C}
The overall optical efficiencies for both system and ancilla modes are given by
\begin{eqnarray}
&&\epsilon_{s}=\epsilon_{Opt}\cdot\epsilon_{channel}\cdot\cdot\epsilon_{SMF}^{2}\cdot\epsilon_{SPAD}
  \\& & 
\epsilon_{a}=\epsilon_{Opt}\cdot\epsilon_{SMF}\cdot\epsilon_{SPAD}\,,
\end{eqnarray}
where the single-photon transmission efficiency of the main optical
components is $\epsilon_{Opt}=0.9$, the single-mode fiber coupling
efficiency is $\epsilon_{SMF}=0.73$ and the SPADs detection efficiency
is $\epsilon_{SPAD}=0.7$. Accordingly the two-photon optical
efficiency was
\begin{equation}
\epsilon_{s,a}=\epsilon_{s}\cdot\epsilon_{a}\approx0.15\cdot\epsilon_{channel}
\end{equation}

The channel optical efficiencies $\epsilon_{channel}$ were $0.6$ for
the AD-channel and $0.98$ for the PD-channel and D-channel. Thus, the
overall coincidence efficiencies $\epsilon_{s,a}$ are 0.09 and 0.15,
respectively.

The propagated standard deviation in our data was calculated from by
Monte Carlo simulations of Poissonian statistics on the photon
coincidence counts. In our experiment the main sources of error were:
\begin{enumerate}
\item The setting of the rotation angle in the LC and HWPs.
\item The setting of the LC voltage for the precise retardance.
\item The statistical propagation due to the mixing of independent Pauli operations.
\end{enumerate}

The first source of error is negligible, because all waveplates
contained in the channels and used for the projective measurements and
the LC were calibrated to a precision of $\pm 0.1$ degrees. For
waveplates at an angle around $0^{\circ}$-degrees or
$45^{\circ}$-degrees, the dependence of counts on a small angle
deviation is linear, thus the overall error is strongly dominated by
the Poissonian statistics on the counts. This is not valid in the
AD-channel configuration, where the interference leads to a strong
non-linearity between waveplate angles and coincidence counts,
governed by $\omega=\frac{\arccos(-\sqrt{1-\eta})}{2}$. Accordingly, we
considered a prudent error ($0.5^{\circ}$-degrees) on the angle,
leading to a considerable increase of the error on the damping
parameter.

The second source of error is negligible as well, because when
performing the calibration of the LC we use horizontally polarized
light, the LC at 45 degrees and a PBS. Here, by scanning the voltage
on the LC we verified a non-linearly retardance, which is almost flat
around either the minimum and maximum voltages used in the experiment,
$V_{low}=0.5V$ and $V_{high}=25.0V$, respectively. We confirmed a
visibility of 0.989 between these two voltages, which represents a
much higher value than the fidelity of the state itself, meaning that
any slight imperfection of $\mathbb{I}$ and $\sigma$ operations will
weakly affect the data tendency compared to the noise coming from the
state generation.

The third source of error is the error propagation that comes from
mixing counts coming from different Pauli operations, suitably
weighted. This error is already included in our Monte Carlo
simulations.

\section{Experimental conditional probabilities}\label{app:D}
In this section we show the tables of all two-photon conditional
probabilities $Q^{(\alpha)}(A|B)$ extracted from the experimental
coincidence measurements.  We show the results for the AD-channel in
Table~\ref{tab:adcTable}, for the PD-channel in Table
\ref{tab:pdcTable}, and for the D-channel in Table \ref{tab:dTable}.
Due to the complement rule for orthogonal input states [for instance,
  $Q^{(\alpha)}(\cdot|H)=1-Q^{(\alpha)}(\cdot|V)$], we only show half
of the total data.

\begin{widetext}

\begin{table}[h!]
\centering
\begin{tabular}{|c|c|c|c|c|c|c|}
 \hline
$\eta$ & $Q^{(z)}(H|H)$ & $Q^{(z)}(V|H)$ & $Q^{(x)}(+|+)$ & $Q^{(x)}(-|+)$ & $Q^{(y)}(L|L)$ & $Q^{(y)}(R|L)$\\
\hline
0 & \text{0.9983 $\pm $ 0.0011} & \text{-0.01396 $\pm $ 0.00078} & \text{1.0001 $\pm $ 0.0011} & \text{-0.0018 $\pm $ 0.0011} & \text{0.0145 $\pm $ 0.0014} & \text{0.9906 $\pm $ 0.0014} \\
0.05 & \text{0.9642 $\pm $ 0.0017} & \text{0.0195 $\pm $ 0.0015} & \text{0.9661 $\pm $ 0.0017} & \text{0.0329 $\pm $ 0.0017} & \text{0.0480 $\pm $ 0.0019} & \text{0.9551 $\pm $ 0.0019} \\
0.1 & \text{0.9300 $\pm $ 0.0021} & \text{0.0530 $\pm $ 0.0020} & \text{0.9322 $\pm $ 0.0021} & \text{0.0674 $\pm $ 0.0021} & \text{0.0815 $\pm $ 0.0023} & \text{0.9199 $\pm $ 0.0023} \\
0.15 & \text{0.8958 $\pm $ 0.0024} & \text{0.0866 $\pm $ 0.0023} & \text{0.8983 $\pm $ 0.0024} & \text{0.1018 $\pm $ 0.0024} & \text{0.1149 $\pm $ 0.0025} & \text{0.8849 $\pm $ 0.0026} \\
0.2 & \text{0.8617 $\pm $ 0.0026} & \text{0.1201 $\pm $ 0.0026} & \text{0.8646 $\pm $ 0.0027} & \text{0.1360 $\pm $ 0.0027} & \text{0.1482 $\pm $ 0.0028} & \text{0.8501 $\pm $ 0.0028} \\
0.25 & \text{0.8275 $\pm $ 0.0028} & \text{0.1537 $\pm $ 0.0028} & \text{0.8311 $\pm $ 0.0029} & \text{0.1700 $\pm $ 0.0029} & \text{0.1814 $\pm $ 0.0030} & \text{0.8156 $\pm $ 0.0030} \\
0.3 & \text{0.7934 $\pm $ 0.0030} & \text{0.1872 $\pm $ 0.0030} & \text{0.7976 $\pm $ 0.0031} & \text{0.2038 $\pm $ 0.0031} & \text{0.2146 $\pm $ 0.0031} & \text{0.7813 $\pm $ 0.0032} \\
0.35 & \text{0.7593 $\pm $ 0.0032} & \text{0.2208 $\pm $ 0.0032} & \text{0.7643 $\pm $ 0.0032} & \text{0.2375 $\pm $ 0.0032} & \text{0.2477 $\pm $ 0.0033} & \text{0.7472 $\pm $ 0.0033} \\
0.4 & \text{0.7252 $\pm $ 0.0033} & \text{0.2543 $\pm $ 0.0033} & \text{0.7310 $\pm $ 0.0034} & \text{0.2709 $\pm $ 0.0034} & \text{0.2806 $\pm $ 0.0034} & \text{0.7133 $\pm $ 0.0034} \\
0.45 & \text{0.6911 $\pm $ 0.0034} & \text{0.2879 $\pm $ 0.0034} & \text{0.6979 $\pm $ 0.0035} & \text{0.3042 $\pm $ 0.0035} & \text{0.3135 $\pm $ 0.0035} & \text{0.6796 $\pm $ 0.0035} \\
0.5 & \text{0.6570 $\pm $ 0.0035} & \text{0.3215 $\pm $ 0.0035} & \text{0.6649 $\pm $ 0.0035} & \text{0.3373 $\pm $ 0.0035} & \text{0.3464 $\pm $ 0.0036} & \text{0.6462 $\pm $ 0.0036} \\
0.55 & \text{0.6230 $\pm $ 0.0035} & \text{0.3552 $\pm $ 0.0036} & \text{0.6321 $\pm $ 0.0036} & \text{0.3702 $\pm $ 0.0036} & \text{0.3791 $\pm $ 0.0036} & \text{0.6130 $\pm $ 0.0037} \\
0.6 & \text{0.5889 $\pm $ 0.0036} & \text{0.3888 $\pm $ 0.0037} & \text{0.5993 $\pm $ 0.0037} & \text{0.4030 $\pm $ 0.0036} & \text{0.4118 $\pm $ 0.0037} & \text{0.5799 $\pm $ 0.0037} \\
\hline
\end{tabular}
\caption{Experimental conditional probabilities in AD-channel.}
\label{tab:adcTable}
\end{table}

\begin{table}[h!]
\centering
\begin{tabular}{|c|c|c|c|c|c|c|}
 \hline
$q$ & $Q^{(z)}(H|H)$ & $Q^{(z)}(V|H)$ & $Q^{(x)}(+|+)$ & $Q^{(x)}(-|+)$ & $Q^{(y)}(L|L)$ & $Q^{(y)}(R|L)$\\
\hline
0 & \text{1.0005 $\pm $ 0.0015} & \text{-0.0163 $\pm $ 0.0014} & \text{1.0023 $\pm $ 0.0015} & \text{-0.0040 $\pm $ 0.0015} & \text{0.0124 $\pm $ 0.0017} & \text{0.9928 $\pm $ 0.0017} \\
0.05 & \text{1.0008 $\pm $ 0.0015} & \text{-0.0162 $\pm $ 0.0014} & \text{0.9766 $\pm $ 0.0019} & \text{0.0214 $\pm $ 0.0019} & \text{0.0375 $\pm $ 0.0020} & \text{0.9664 $\pm $ 0.0020} \\
0.1 & \text{1.0012 $\pm $ 0.0015} & \text{-0.0161 $\pm $ 0.0014} & \text{0.9509 $\pm $ 0.0022} & \text{0.0469 $\pm $ 0.0022} & \text{0.0625 $\pm $ 0.0023} & \text{0.9402 $\pm $ 0.0023} \\
0.15 & \text{1.0016 $\pm $ 0.0015} & \text{-0.0160 $\pm $ 0.0014} & \text{0.9252 $\pm $ 0.0024} & \text{0.0723 $\pm $ 0.0024} & \text{0.0875 $\pm $ 0.0025} & \text{0.9140 $\pm $ 0.0025} \\
0.2 & \text{1.0019 $\pm $ 0.0015} & \text{-0.0159 $\pm $ 0.0014} & \text{0.8996 $\pm $ 0.0026} & \text{0.0977 $\pm $ 0.0026} & \text{0.1124 $\pm $ 0.0027} & \text{0.8880 $\pm $ 0.0027} \\
0.25 & \text{1.0023 $\pm $ 0.0015} & \text{-0.0159 $\pm $ 0.0014} & \text{0.8740 $\pm $ 0.0028} & \text{0.1232 $\pm $ 0.0028} & \text{0.1374 $\pm $ 0.0028} & \text{0.8620 $\pm $ 0.0029} \\
0.3 & \text{1.0027 $\pm $ 0.0015} & \text{-0.0158 $\pm $ 0.0014} & \text{0.8484 $\pm $ 0.0029} & \text{0.1485 $\pm $ 0.0029} & \text{0.1623 $\pm $ 0.0030} & \text{0.8361 $\pm $ 0.0030} \\
0.35 & \text{1.0030 $\pm $ 0.0015} & \text{-0.0157 $\pm $ 0.0014} & \text{0.8228 $\pm $ 0.0030} & \text{0.1739 $\pm $ 0.0030} & \text{0.1871 $\pm $ 0.0031} & \text{0.8103 $\pm $ 0.0031} \\
0.4 & \text{1.0034 $\pm $ 0.0015} & \text{-0.0156 $\pm $ 0.0014} & \text{0.7973 $\pm $ 0.0032} & \text{0.1993 $\pm $ 0.0032} & \text{0.2120 $\pm $ 0.0032} & \text{0.7846 $\pm $ 0.0033} \\
0.45 & \text{1.0038 $\pm $ 0.0015} & \text{-0.0155 $\pm $ 0.0014} & \text{0.7719 $\pm $ 0.0033} & \text{0.2246 $\pm $ 0.0033} & \text{0.2368 $\pm $ 0.0033} & \text{0.7590 $\pm $ 0.0034} \\
0.5 & \text{1.0042 $\pm $ 0.0015} & \text{-0.0154 $\pm $ 0.0014} & \text{0.7464 $\pm $ 0.0034} & \text{0.2499 $\pm $ 0.0034} & \text{0.2616 $\pm $ 0.0034} & \text{0.7334 $\pm $ 0.0034} \\
0.55 & \text{1.0045 $\pm $ 0.0015} & \text{-0.0153 $\pm $ 0.0014} & \text{0.7210 $\pm $ 0.0034} & \text{0.2752 $\pm $ 0.0034} & \text{0.2864 $\pm $ 0.0035} & \text{0.7080 $\pm $ 0.0035} \\
0.6 & \text{1.0049 $\pm $ 0.0015} & \text{-0.0152 $\pm $ 0.0014} & \text{0.6956 $\pm $ 0.0035} & \text{0.3005 $\pm $ 0.0035} & \text{0.3111 $\pm $ 0.0035} & \text{0.6826 $\pm $ 0.0036} \\
0.65 & \text{1.0052 $\pm $ 0.0015} & \text{-0.0151 $\pm $ 0.0014} & \text{0.6702 $\pm $ 0.0036} & \text{0.3257 $\pm $ 0.0036} & \text{0.3359 $\pm $ 0.0036} & \text{0.6573 $\pm $ 0.0036} \\
0.7 & \text{1.0056 $\pm $ 0.0015} & \text{-0.0150 $\pm $ 0.0014} & \text{0.6449 $\pm $ 0.0036} & \text{0.3510 $\pm $ 0.0036} & \text{0.3605 $\pm $ 0.0036} & \text{0.6321 $\pm $ 0.0037} \\
0.75 & \text{1.0060 $\pm $ 0.0015} & \text{-0.0150 $\pm $ 0.0014} & \text{0.6196 $\pm $ 0.0037} & \text{0.3762 $\pm $ 0.0037} & \text{0.3852 $\pm $ 0.0037} & \text{0.6070 $\pm $ 0.0037} \\
0.8 & \text{1.0063 $\pm $ 0.0015} & \text{-0.0149 $\pm $ 0.0014} & \text{0.5944 $\pm $ 0.0037} & \text{0.4013 $\pm $ 0.0037} & \text{0.4098 $\pm $ 0.0037} & \text{0.5819 $\pm $ 0.0037} \\
0.85 & \text{1.0067 $\pm $ 0.0015} & \text{-0.0148 $\pm $ 0.0014} & \text{0.5691 $\pm $ 0.0037} & \text{0.4265 $\pm $ 0.0037} & \text{0.4345 $\pm $ 0.0037} & \text{0.5570 $\pm $ 0.0037} \\
0.9 & \text{1.0071 $\pm $ 0.0015} & \text{-0.0147 $\pm $ 0.0014} & \text{0.5439 $\pm $ 0.0037} & \text{0.4517 $\pm $ 0.0038} & \text{0.4590 $\pm $ 0.0037} & \text{0.5321 $\pm $ 0.0038} \\
0.95 & \text{1.0075 $\pm $ 0.0014} & \text{-0.0146 $\pm $ 0.0014} & \text{0.5187 $\pm $ 0.0037} & \text{0.4768 $\pm $ 0.0038} & \text{0.4836 $\pm $ 0.0037} & \text{0.5073 $\pm $ 0.0038} \\
1 & \text{1.0079 $\pm $ 0.0014} & \text{-0.0145 $\pm $ 0.0014} & \text{0.4936 $\pm $ 0.0037} & \text{0.5019 $\pm $ 0.0038} & \text{0.5081 $\pm $ 0.0037} & \text{0.4826 $\pm $ 0.0038} \\
\hline
\end{tabular}
\caption{Experimental conditional probabilities in PD-channel, from
  the family of P-channels.}
\label{tab:pdcTable}
\end{table}

\begin{table}[h!]
\centering
\begin{tabular}{|c|c|c|c|c|c|c|}
 \hline
$q$ & $Q^{(z)}(H|H)$ & $Q^{(z)}(V|H)$ & $Q^{(x)}(+|+)$ & $Q^{(x)}(-|+)$ & $Q^{(y)}(L|L)$ & $Q^{(y)}(R|L)$\\
\hline
0 & \text{0.9983 $\pm $ 0.0011} & \text{-0.01396 $\pm $ 0.00078} & \text{1.0001 $\pm $ 0.0011} & \text{-0.0018 $\pm $ 0.0011} & \text{0.0145 $\pm $ 0.0014} & \text{0.9906 $\pm $ 0.0014} \\
0.017 & \text{0.9642 $\pm $ 0.0017} & \text{0.0195 $\pm $ 0.0015} & \text{0.9661 $\pm $ 0.0017} & \text{0.0329 $\pm $ 0.0017} & \text{0.0480 $\pm $ 0.0019} & \text{0.9551 $\pm $ 0.0019} \\
0.033 & \text{0.9300 $\pm $ 0.0021} & \text{0.0530 $\pm $ 0.0020} & \text{0.9322 $\pm $ 0.0021} & \text{0.0674 $\pm $ 0.0021} & \text{0.0815 $\pm $ 0.0023} & \text{0.9199 $\pm $ 0.0023} \\
0.05 & \text{0.8958 $\pm $ 0.0024} & \text{0.0866 $\pm $ 0.0023} & \text{0.8983 $\pm $ 0.0024} & \text{0.1018 $\pm $ 0.0024} & \text{0.1149 $\pm $ 0.0025} & \text{0.8849 $\pm $ 0.0026} \\
0.067 & \text{0.8617 $\pm $ 0.0026} & \text{0.1201 $\pm $ 0.0026} & \text{0.8646 $\pm $ 0.0027} & \text{0.1360 $\pm $ 0.0027} & \text{0.1482 $\pm $ 0.0028} & \text{0.8501 $\pm $ 0.0028} \\
0.083 & \text{0.8275 $\pm $ 0.0028} & \text{0.1537 $\pm $ 0.0028} & \text{0.8311 $\pm $ 0.0029} & \text{0.1700 $\pm $ 0.0029} & \text{0.1814 $\pm $ 0.0030} & \text{0.8156 $\pm $ 0.0030} \\
0.1 & \text{0.7934 $\pm $ 0.0030} & \text{0.1872 $\pm $ 0.0030} & \text{0.7976 $\pm $ 0.0031} & \text{0.2038 $\pm $ 0.0031} & \text{0.2146 $\pm $ 0.0031} & \text{0.7813 $\pm $ 0.0032} \\
0.117 & \text{0.7593 $\pm $ 0.0032} & \text{0.2208 $\pm $ 0.0032} & \text{0.7643 $\pm $ 0.0032} & \text{0.2375 $\pm $ 0.0032} & \text{0.2477 $\pm $ 0.0033} & \text{0.7472 $\pm $ 0.0033} \\
0.133 & \text{0.7252 $\pm $ 0.0033} & \text{0.2543 $\pm $ 0.0033} & \text{0.7310 $\pm $ 0.0034} & \text{0.2709 $\pm $ 0.0034} & \text{0.2806 $\pm $ 0.0034} & \text{0.7133 $\pm $ 0.0034} \\
0.15 & \text{0.6911 $\pm $ 0.0034} & \text{0.2879 $\pm $ 0.0034} & \text{0.6979 $\pm $ 0.0035} & \text{0.3042 $\pm $ 0.0035} & \text{0.3135 $\pm $ 0.0035} & \text{0.6796 $\pm $ 0.0035} \\
0.167 & \text{0.6570 $\pm $ 0.0035} & \text{0.3215 $\pm $ 0.0035} & \text{0.6649 $\pm $ 0.0035} & \text{0.3373 $\pm $ 0.0035} & \text{0.3464 $\pm $ 0.0036} & \text{0.6462 $\pm $ 0.0036} \\
0.183 & \text{0.6230 $\pm $ 0.0035} & \text{0.3552 $\pm $ 0.0036} & \text{0.6321 $\pm $ 0.0036} & \text{0.3702 $\pm $ 0.0036} & \text{0.3791 $\pm $ 0.0036} & \text{0.6130 $\pm $ 0.0037} \\
0.2 & \text{0.5889 $\pm $ 0.0036} & \text{0.3888 $\pm $ 0.0037} & \text{0.5993 $\pm $ 0.0037} & \text{0.4030 $\pm $ 0.0036} & \text{0.4118 $\pm $ 0.0037} & \text{0.5799 $\pm $ 0.0037} \\
0.217 & \text{0.5549 $\pm $ 0.0036} & \text{0.4224 $\pm $ 0.0037} & \text{0.5666 $\pm $ 0.0037} & \text{0.4356 $\pm $ 0.0037} & \text{0.4444 $\pm $ 0.0037} & \text{0.5472 $\pm $ 0.0037} \\
0.233 & \text{0.5209 $\pm $ 0.0036} & \text{0.4561 $\pm $ 0.0037} & \text{0.5341 $\pm $ 0.0037} & \text{0.4680 $\pm $ 0.0037} & \text{0.4769 $\pm $ 0.0037} & \text{0.5145 $\pm $ 0.0037} \\
0.25 & \text{0.4869 $\pm $ 0.0036} & \text{0.4898 $\pm $ 0.0038} & \text{0.5016 $\pm $ 0.0037} & \text{0.5003 $\pm $ 0.0037} & \text{0.5093 $\pm $ 0.0037} & \text{0.4822 $\pm $ 0.0037} \\
0.267 & \text{0.4529 $\pm $ 0.0036} & \text{0.5234 $\pm $ 0.0038} & \text{0.4693 $\pm $ 0.0037} & \text{0.5324 $\pm $ 0.0037} & \text{0.5417 $\pm $ 0.0037} & \text{0.4500 $\pm $ 0.0037} \\
0.283 & \text{0.4189 $\pm $ 0.0036} & \text{0.5571 $\pm $ 0.0037} & \text{0.4371 $\pm $ 0.0037} & \text{0.5643 $\pm $ 0.0037} & \text{0.5740 $\pm $ 0.0037} & \text{0.4180 $\pm $ 0.0037} \\
0.3 & \text{0.3849 $\pm $ 0.0036} & \text{0.5908 $\pm $ 0.0037} & \text{0.4050 $\pm $ 0.0036} & \text{0.5961 $\pm $ 0.0036} & \text{0.6062 $\pm $ 0.0036} & \text{0.3862 $\pm $ 0.0036} \\
0.317 & \text{0.3509 $\pm $ 0.0035} & \text{0.6245 $\pm $ 0.0037} & \text{0.3730 $\pm $ 0.0036} & \text{0.6276 $\pm $ 0.0036} & \text{0.6383 $\pm $ 0.0036} & \text{0.3546 $\pm $ 0.0036} \\
0.333 & \text{0.3169 $\pm $ 0.0034} & \text{0.6583 $\pm $ 0.0036} & \text{0.3411 $\pm $ 0.0035} & \text{0.6591 $\pm $ 0.0035} & \text{0.6703 $\pm $ 0.0035} & \text{0.3232 $\pm $ 0.0035} \\
\hline
\end{tabular}
\caption{Experimental conditional probabilities in D-channel, from the
  family of P-channels.}
\label{tab:dTable}
\end{table}
\end{widetext}

\end{document}